\documentclass[
reprint,           
superscriptaddress,
amsmath,           
amssymb,           
aps,               
prl,               
notitlepage,       
longbibliography,  
floatfix,          
nofootinbib,
]{revtex4-1}

\usepackage{tensor}     
\usepackage{graphicx}   
\usepackage[
colorlinks=true,        
citecolor=blue,         
linkcolor=blue,         
urlcolor=blue           
]{hyperref}             
\usepackage{bm}         
\usepackage{xcolor}     
\usepackage{lipsum}
\usepackage{color}      
\usepackage[utf8]{inputenc} 
\usepackage[section]{placeins} 
\usepackage{multirow}

\newcommand{\nc}{\newcommand*} 

\nc{\al}{\alpha}
\nc{\s}{\sigma}
\nc{\kp}{\kappa}
\nc{\dt}{\delta}
\nc{\Dt}{\Delta}
\nc{\Ld}{\Lambda}
\nc{\p}{\partial}
\nc{\Gm}{\Gamma}
\nc{\om}{\omega}
\nc{\Om}{\Omega}
\nc{\rd}{\mathrm{d}}
\nc{\Od}{\mathcal{O}} 
\def\({\left(}
\def\){\right)}
\def\[{\left[}
\def\]{\right]}
\def\e{\begin{equation}}
\def\q{\end{equation}}
\def\m{\begin{eqnarray}}
\def\n{\end{eqnarray}}
\nc{\Eq}[1]{Eq.~\eqref{#1}}     
\nc{\Fig}[1]{Fig.~\ref{#1}}     
\nc{\Table}[1]{Table~\ref{#1}}  
\nc{\Sec}[1]{Sec.~\ref{#1}}     
\nc{\Msun}{M_\odot}             
\nc{\fpbh}{f_{\mathrm{PBH}}}    
\nc{\fpbhn}{f_{\mathrm{pbh0}}}    
\nc{\mR}{\mathcal{R}} 
\nc{\seq}{\sigma_{\mathrm{eq}}}
\nc{\ogw}{\Omega_{\mathrm{GW}}}
\nc{\gpcyr}{\mathrm{Gpc}^{-3}\,\mathrm{yr}^{-1}}
\nc{\lvc}{LIGO-Virgo} 
\nc{\SNR}{\mathrm{SNR}} 
\nc{\mmin}{{m_{\mathrm{min}}}}
\nc{\mmax}{{m_{\mathrm{max}}}}
\nc{\Mmin}{{M_{\mathrm{min}}}}
\nc{\fmin}{{f_{\mathrm{min}}}}
\nc{\VT}{\mathrm{VT}}
\nc{\rhoGW}{\rho_{\mathrm{GW}}}
\nc{\vth}{\vec{\theta}}
\nc{\vd}{\vec{d}}
\nc{\vla}{\vec{\lambda}}
\nc{\Nobs}{N_{\mathrm{obs}}}
\nc{\av}[1]{\langle #1 \rangle} 
\nc{\km}{\mathrm{km}}
\nc{\Mpc}{\mathrm{Mpc}}
\nc{\Tobs}{T_{\mathrm{obs}}}
\nc{\Ntemp}{N_{\mathrm{temp}}}
\nc{\fyr}{f_{\mathrm{yr}}}
\nc{\addref}{[\textcolor{red}{add ref}] } 
\nc{\eg}{\textit{e.g.~}}
\nc{\app}{\approx}
\nc{\hf}{\frac{1}{2}}
\nc{\discuss}{\textcolor{red}{Add discussion here!}}
\nc{\red}[1]{\textcolor{red}{#1}}
\nc{\hp}{h_+} 
\nc{\hc}{h_{\times}} 
\nc{\Oh}{\hat{\Omega}}
\nc{\vx}{\vec{x}}
\nc{\mh}{\hat{m}}
\nc{\nh}{\hat{n}}
\nc{\zh}{\hat{z}}
\nc{\ph}{\hat{p}}
\nc{\A}[1]{\mathcal{A}_{#1}}
\nc{\Ogw}[1]{\Omega_{\mathrm{#1}}}
\nc{\bn}[1]{\dt\bm{t}_{\text{#1}}}
\nc{\bC}[1]{\bm{C}_{\text{#1}}}
\nc{\NTOA}{N_{\text{TOA}}}
\nc{\Nmode}{{N_{\text{mode}}}}
\nc{\ARN}{A_{\rm{RN}}}
\nc{\gRN}{\gamma_{\rm{RN}}}
\nc{\bS}{\mathbf{\Sigma}}
\nc{\br}{\mathbf{r}}
\nc{\bN}{\mathbf{R}}
\nc{\Agw}{A_\mathrm{GWB}}
\nc{\UCP}{\mathrm{UCP}}
\nc{\TT}{\mathrm{TT}}
\nc{\ST}{\mathrm{ST}}
\nc{\SL}{\mathrm{SL}}
\nc{\VL}{\mathrm{VL}}
\nc{\BFST}{$107 \pm 7$}

\begin{document}
	
\title{Confronting the primordial black hole scenario with the gravitational-wave events detected by LIGO-Virgo}
	
\author{Zu-Cheng Chen}
\email{chenzucheng@itp.ac.cn} 
\affiliation{CAS Key Laboratory of Theoretical Physics, 
		Institute of Theoretical Physics, Chinese Academy of Sciences,
		Beijing 100190, China}
\affiliation{School of Physical Sciences, 
		University of Chinese Academy of Sciences, 
		No. 19A Yuquan Road, Beijing 100049, China}
	
\author{Chen Yuan}
\email{yuanchen@itp.ac.cn}
\affiliation{CAS Key Laboratory of Theoretical Physics, 
		Institute of Theoretical Physics, Chinese Academy of Sciences,
		Beijing 100190, China}
\affiliation{School of Physical Sciences, 
		University of Chinese Academy of Sciences, 
		No. 19A Yuquan Road, Beijing 100049, China}
	
\author{Qing-Guo Huang}
\email{Corresponding author: huangqg@itp.ac.cn}
\affiliation{CAS Key Laboratory of Theoretical Physics, 
		Institute of Theoretical Physics, Chinese Academy of Sciences,
		Beijing 100190, China}
\affiliation{School of Physical Sciences, 
		University of Chinese Academy of Sciences, 
		No. 19A Yuquan Road, Beijing 100049, China}
\affiliation{School of Fundamental Physics and Mathematical Sciences
		Hangzhou Institute for Advanced Study, UCAS, Hangzhou 310024, China}
\affiliation{Center for Gravitation and Cosmology, 
		College of Physical Science and Technology, 
		Yangzhou University, Yangzhou 225009, China}

\date{\today}
	
\begin{abstract}

Adopting a binned method, we model-independently reconstruct the mass function of primordial black holes (PBHs) from GWTC-3 and find that such a PBH mass function can be explained by a broad red-tilted power spectrum of curvature perturbations. Even though GW190521 with component masses in upper mass gap $(m>65M_\odot)$ can be naturally interpreted in the PBH scenario, the events (including GW190814, GW190425, GW200105, and GW200115) with component masses in the light mass range $(m<3M_\odot)$ are quite unlikely to be explained by binary PBHs although there are no electromagnetic counterparts because the corresponding PBH merger rates are much smaller than those given by LIGO-Virgo. Furthermore, we predict that both the gravitational-wave (GW) background generated by the binary PBHs and the scalar-induced GWs accompanying the formation of PBHs should be detected by the ground-based and space-borne GW detectors and pulsar timing arrays in the future.

\end{abstract}

\maketitle
	
\textit{Introduction.} Primordial black holes (PBHs) \cite{Carr:1974nx,Carr:1975qj}  are formed in the very early Universe due to the collapse of over-densed regions which are generated by the enhanced curvature power spectrum on small scales compared to those on the cosmic microwave background (CMB) scales. PBHs can not only represent the dark matter (DM) in the Universe, but also may potentially provide an explanation to the merger events detected by LIGO-Virgo Collaboration (LVC) if the fraction of the stellar mass PBHs in cold dark matter (CDM) is $\fpbh\sim\mathrm{few}\times 10^{-3}$ \cite{Sasaki:2016jop,Chen:2018czv,Raidal:2018bbj,DeLuca:2020qqa,Hall:2020daa,Bhagwat:2020bzh,Hutsi:2020sol,Wong:2020yig,DeLuca:2021wjr,Franciolini:2021tla}.

Among all the merger events detected by LVC, some of them are likely to have ambiguities in the astrophysical scenario. Firstly, the primary component of GW190521 has a high probability to be within the pair-instability supernovae mass gap \cite{LIGOScientific:2020iuh}, implying that the primary black hole (BH) might not have a stellar origin. Secondly, even though the component masses of GW190425 lie in the mass range of $[1.12\Msun, 2.52\Msun]$ and are consistent with the individual binary component being neutron star (NS) \cite{LIGOScientific:2020aai}, the source-frame chirp mass and total mass are significantly larger than any known binary NS system. Thirdly, GW190814 is reported to have a compact object with a mass of $2.5-2.67\Msun$ \cite{LIGOScientific:2020zkf}, which falls in the ``lower mass gap'' where no NSs or BHs have been observed in a double compact-object system. Finally, LVC recently reported GW200105 and GW200115 \cite{LIGOScientific:2021qlt} in which the secondary component masses are respectively $1.9_{-0.2}^{+0.3}\Msun$ and $1.5_{-0.3}^{+0.7}\Msun$, indicating that both of them are well below the maximal mass of an NS. In addition, there are no electromagnetic counterparts to confirm and PBHs are speculated to provide an explanation to these five gravitational-wave (GW) events \cite{LIGOScientific:2020aai,LIGOScientific:2020zkf,LIGOScientific:2021qlt,DeLuca:2020sae}.

In this letter, we will give a comprehensive investigation for the possibility that if the GW events, in particular the five events mentioned above, can be explained in the PBH scenario. We adopt a binned method to model-independently reconstruct the PBH mass function from GWTC-3 \cite{LIGOScientific:2021djp}, and we find that GW190521 can be explained by a binary PBH merger. But the other four GW events (GW190814, GW190425, GW200105, and GW200115) are quite unlikely to be interpreted as binary PBHs because the corresponding merger rates of binary PBHs are much smaller than those given by LVC even though the PBH mass function around $\sim 1M_\odot$ is significantly enhanced due to the softening of the equation of state during the QCD phase transition \cite{Byrnes:2018clq}. 

In addition, we compute the gravitational-wave background (GWB) from both the PBH binary coalescences \cite{Chen:2018rzo} and the so-called scalar-induced gravitational waves (SIGWs) generated by the curvature perturbation during the formation of PBHs \cite{Saito:2008jc}. We find that the GWB associated with PBHs is compatible with the current limits of observations and should be detected by the ground-based and space-borne GW detectors and pulsar timing arrays in the future.

{\it A model-independent reconstruction of PBH mass function.} In this work, the PBH mass is considered in the range of $[1M_\odot,130M_\odot]$. To infer the PBH mass function from the GWTC-3, we adopt a model-independent approach using the following binned mass function 
\e\label{para} 
P(m) = \begin{cases} 
    P_1, & 1\, \Msun \leq m < 3\, \Msun \\
    P_2, & 3\, \Msun \leq m < 10\, \Msun \\
    P_3, & 10\, \Msun \leq m < 40\, \Msun \\
    P_4, & 40\, \Msun \leq m < 80\, \Msun \\
    P_5, & 80\, \Msun \leq m \leq 130\, \Msun
\end{cases}
\q 
in which the mass function $P(m)$ is normalized by $\int P(m) dm = 1$. Therefore, only four of $P_i (i=1, \cdots, 5)$ are independent, and $\vth = \{P_1, P_2, P_3, P_4\}$ are chosen to be the free parameters. The merger rate density in units of $\gpcyr$ for a general mass function, $P(m|\vth)$, takes the form of \cite{Chen:2018czv}
\m\label{calR} 
\mR_{12}&&(t|\vth) \app 2.8 \cdot 10^6 \times \({\frac{t}{t_0}}\)^{-\frac{34}{37}} \fpbh^2 (0.7\fpbh^2+\sigma_{\mathrm{eq}}^2)^{-{21\over 74}} \nonumber \\
&& \times  \min\(\frac{P(m_1|\vth)}{m_1}, \frac{P(m_2|\vth)}{m_2}\) \({P(m_1|\vth)\over m_1}+{P(m_2|\vth)\over m_2}\) \nonumber \\
&& \times (m_1 m_2)^{{3\over 37}} (m_1+m_2)^{36\over 37},
\n
where the component masses $m_1$ and $m_2$ are in units of $\Msun$, $\fpbh \equiv \Omega_{\mathrm{PBH}}/\Omega_{\mathrm{CDM}}$ is the energy density fraction of PBHs in CDM, and $\sigma_{\mathrm{eq}}\approx 0.005$ \cite{Ali-Haimoud:2017rtz,Chen:2018czv} is the variance of density perturbations of the rest CDM on scale of order $\Od(10^0\sim10^3) M_\odot$ at radiation-matter equality. 

We perform the hierarchical Bayesian inference \cite{LIGOScientific:2016kwr,LIGOScientific:2016ebi,TheLIGOScientific:2016pea,Wysocki:2018mpo,Fishbach:2018edt,Mandel:2018mve,Thrane:2018qnx} to extract the population parameters $\{\vth, \fpbh\}$ from observed BBHs. Given the data of $N$ binary BH (BBH) detections, $\vd = (d_1, \dots, d_N)$, the likelihood for an inhomogeneous Poisson process is \cite{Wysocki:2018mpo,Fishbach:2018edt,Mandel:2018mve,Thrane:2018qnx}
\e\label{likelihood}
p(\vd|\vth, R) \propto e^{-\beta(\vth)} \prod_i^N 
\int \rd\vla\ p(\vla|d_i) \ \mR_{12}(\vla|\vth),
\q 
where $\vla \equiv \{m_1, m_2\}$, $p(\vla|d_i)$ is the posterior of an individual event, and $\beta(\vth) \equiv \int \rd\vla\ VT(\vla)\ \mR_{12}(\vla|\vth)$ where $VT(\vla)$ is the spacetime sensitivity volume of \lvc\ detectors. We use the GW events from GWTC-3 by discarding events with false alarm rate larger than $1\ \mathrm{yr}^{-1}$, 
and events with the secondary component mass smaller than $3\Msun$ to avoid contamination from putative events involving neutron stars \cite{DeLuca:2021wjr}. On the other hand, although PBHs are expected to have negligible spin at formation \cite{DeLuca:2019buf,Mirbabayi:2019uph}, they might become fast rotating through accretion effects \cite{DeLuca:2020bjf,DeLuca:2020fpg}. However, given that the accretion model is very sensitive to the cut-off points of the red-shift and there is no evidence to support the accretion effects on PBHs so far, we consider two cases where in \textbf{case I} we discard the events with non-vanishing effective spin while we keep these events in \textbf{case II}.

\begin{figure}[htbp!]
    \centering
    \includegraphics[width = 0.48\textwidth]{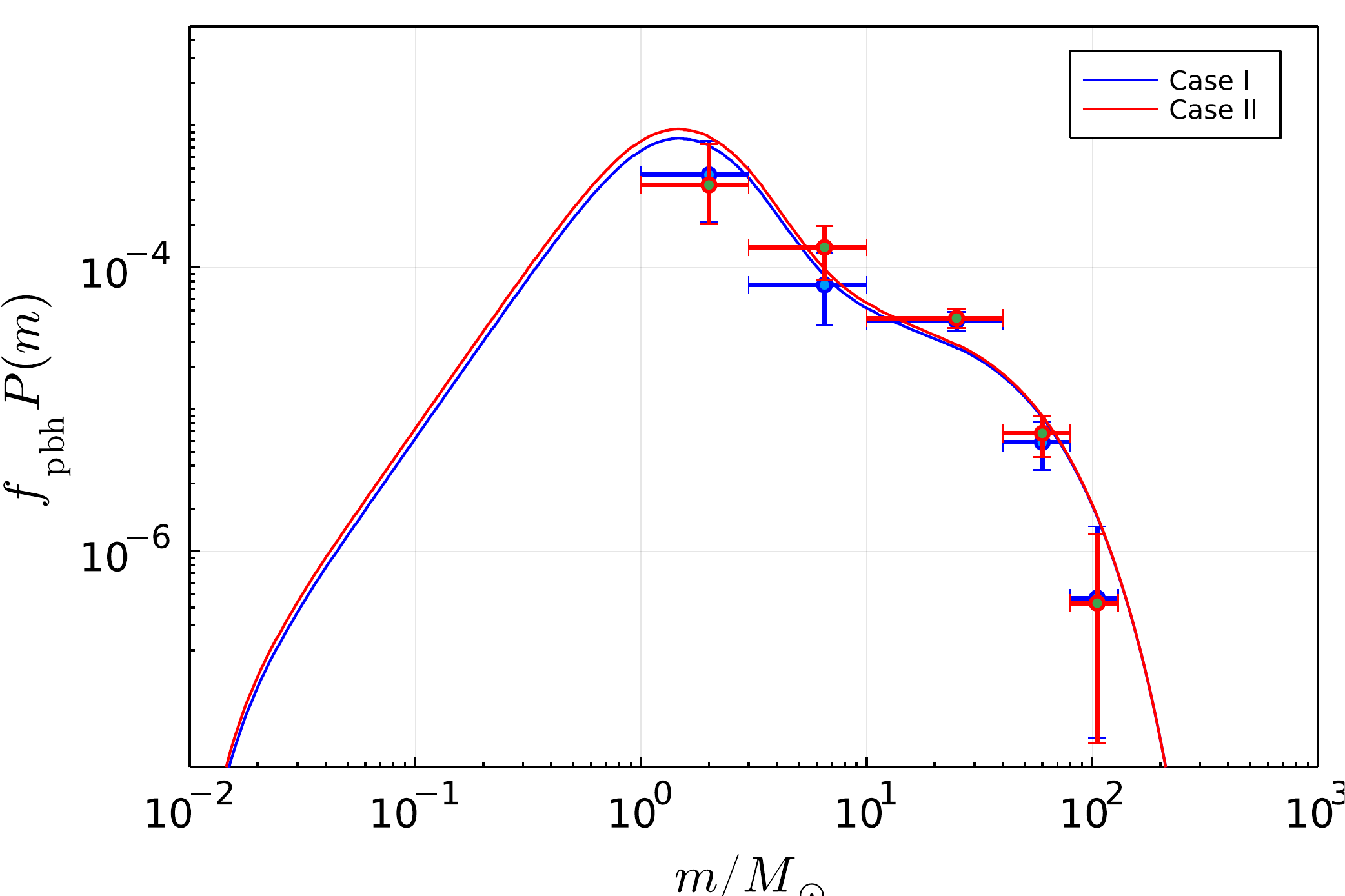}
    \caption{\label{fpbh_Pm_post}
        The median values and the $90\%$ credible intervals of the population parameters $\fpbh P_i$ with $i=1, \cdots, 5$. The blue (red) curves correspond to the mass function of PBHs generated by a broad tilted curvature power spectrum corresponding to discarding (keeping) non-vanishing effective spin events.
    }
\end{figure}

The median value and $90\%$ equal-tailed credible intervals for the parameters $\{\vth, \fpbh\}$ are represented by crosses in \Fig{fpbh_Pm_post}. For the case I, the results are 
$P_1 = 1.5^{+0.6}_{-0.7} \times 10^{-1} \Msun^{-1}$,
$P_2 = 2.5^{+1.8}_{-1.1} \times 10^{-2} \Msun^{-1}$,
$P_3 = 1.4^{+0.4}_{-0.3} \times 10^{-2} \Msun^{-1}$,
$P_4 = 2.0^{+0.9}_{-0.7} \times 10^{-3} \Msun^{-1}$,
and $\fpbh = 2.9^{+0.7}_{-0.6}\times 10^{-3}$. 
Therefore, the total local merger rate is $154^{+186}_{-88}\, \gpcyr$ according to Eq.~(\ref{calR}). 
For the case II, the results are 
$P_1 = 1.1^{+0.8}_{-0.5} \times 10^{-1} \Msun^{-1}$,
$P_2 = 4.2^{+1.2}_{-1.8} \times 10^{-2} \Msun^{-1}$,
$P_3 = 1.3^{+0.3}_{-0.3} \times 10^{-2} \Msun^{-1}$,
$P_4 = 2.0^{+0.7}_{-0.6} \times 10^{-3} \Msun^{-1}$,
and $\fpbh = 3.4^{+0.7}_{-0.5}\times 10^{-3}$. 
Therefore, the total local merger rate is $175^{+144}_{-68}\, \gpcyr$.

From our binned reconstruction of PBH mass function, the inferred merger rates of binary PBHs for the events with component masses in the upper mass gap $(m>65M_\odot)$ and the light mass range $(m<3M_\odot)$ are listed in the third column of Tab.~\ref{merger_rate}. Compared to the merger rates given by LVC (the second column of Tab.~\ref{merger_rate}), we conclude that the upper mass gap event GW190521 can be explained by the PBH scenario, while the events with at least one component mass being smaller than $3\Msun$ (GW190425, GW190814, GW200105, and GW200115) are quite unlikely to be explained in the PBH scenario because the merger rates of corresponding binary PBHs are at least one order of magnitude smaller than those given by LVC.

\begin{table}[htbp!]
    \begin{center}
        \begin{tabular}{c|c|cc}
            \hline\hline
            \multirow{2}{*}{Event} & \multirow{2}{*}{$R_\mathrm{LVC}[\gpcyr]$} & \multicolumn{2}{c}{$R_\mathrm{PBH}[\gpcyr]$}\\
                  & & case I & case II \\
            \hline
            GW190521  & $0.13^{+0.30}_{-0.11}$ \cite{LIGOScientific:2020iuh} & $0.12^{+0.11}_{-0.07}$ & $0.16^{+0.11}_{-0.08}$\\
            \hline
            GW190425 & $460^{+1050}_{-390}$ \cite{LIGOScientific:2020aai} & $3.6^{+6.8}_{-2.8}$ & $2.5^{+6.8}_{-1.8}$\\
            \hline
            GW190814 & $7^{+16}_{-6}$ \cite{LIGOScientific:2020zkf} & $0.13^{+0.09}_{-0.07}$ & $0.12^{+0.11}_{-0.05}$\\
            \hline
            GW200105 & $16^{+38}_{-14}$ \cite{LIGOScientific:2021qlt} & $1.9^{+1.8}_{-1.0}$ & $2.8^{+2.1}_{-1.4}$\\
            \hline
            GW200115 & $36^{+82}_{-30}$ \cite{LIGOScientific:2021qlt} & $6.3^{+7.4}_{-3.8}$ & $8.7^{+8.1}_{-4.5}$\\
            \hline

            \hline
        \end{tabular}
    \end{center}  
    \caption{\label{merger_rate}The local merger rate inferred by \lvc\ and PBH.}
\end{table}

It is also worthy figuring out a physical explanation for the PBH mass function reconstucted from GWTC-3 in Fig.~\ref{fpbh_Pm_post}.  Here, we take into account a broad tilted power spectrum for the curvature perturbations, namely 
\e
\mathcal{P}_\mathcal{R}(k)= A (k/k_{\min})^\alpha\Theta(k-k_{\min})\Theta(k_{\max}-k), 
\label{pzeta}
\q
where $\Theta$ is the heaviside theta function, $k_{\min}=3.2\times10^5\mathrm{Mpc}^{-1}$ and $k_{\max}=9.5\times10^6\mathrm{Mpc}^{-1}$.
After PBHs are formed, they evolve like dust-like matter during radiation dominated era, and then 
\begin{equation}
	\fpbh P(m)=\frac{1}{\Omega_{\mathrm{CDM}}m}\left(\frac{M_{\mathrm{eq}}}{m}\right)^{1 / 2} \beta\left(m\right),
\end{equation}
where $M_{\mathrm{eq}}\approx2.8\times10^{17}\Msun$ is the horizon mass at matter-radiation equality and $\beta(m)$ describes the mass fraction of the Universe that collapse to form PBHs. 
Using Press-Schechter formalism \cite{Press:1973iz}, $\beta(m)$ can be evaluated by integrating the probability distribution function (PDF) $P(\delta)$ of the density contrast $\delta$ that is larger than the threshold value, namely
\e
\beta(m)=\int_{\delta_c}^{\infty}d \delta \frac{m}{M_H}P(\delta),
\q
where 
\e
M_H\simeq 17\left(\frac{g}{10.75}\right)^{-1 / 6}\left(\frac{k}{10^{6} \mathrm{Mpc}^{-1}}\right)^{-2} M_{\odot}
\q
is the horizon mass. Here $g$ is the degress of freedom of relativistic particles, $\delta_c$ is the threshold value of density contrast for the formation of PBHs, $P(\delta)=e^{-\delta^2/(2\sigma_k^2)}/\sqrt{2\pi\sigma_k^2}$ with variance $\sigma_k$ related to the curvature power spectrum by 
\e
\sigma^{2}_k=\(\frac{4}{9}\)^{2}\int_{0}^{\infty} \frac{d q}{q} {W}^{2}(q,R_H)\left(\frac{q}{k}\right)^{4} T^{2}(q,R_H) \mathcal{P}_{\mathcal{R}}(q),
\q
where 
\e
T(k, R_{H})=3 \frac{\sin (k R_{H} / \sqrt{3})-(k R_{H} / \sqrt{3}) \cos (k R_{H} / \sqrt{3})}{(k R_{H} / \sqrt{3})^{3}} 
\q
is the transfer function during radiation dominated era.
A window function $W(k, R_{H})$ is needed to smooth out the density contrast on a comoving length $R\sim k^{-1}$, for which we use a top-hap window function in real space, namely 
\e
W(k, R_{H})=3 \frac{\sin(k R_{H})-(k R_{H}) \cos(k R_{H})}{(k R_{H})^{3}}.
\q
The PBH mass $m$ is related to the density contrast by the critical collapse, namely $m=M_{H} \kappa\left(\delta_m-\delta_{c}\right)^{\gamma}$ \cite{Choptuik:1992jv,Evans:1994pj,Niemeyer:1997mt} with $\kappa=3.3$ and $\gamma=0.36$ \cite{Koike:1995jm}. The nonlinear relation between the density contrast and the curvature perturbation $\delta_\zeta$ leads to $\delta_m=\delta_\zeta-3/8\delta_\zeta^2$ \cite{Young:2019yug,DeLuca:2019qsy,Kawasaki:2019mbl}.
For the equation of state $w=1/3$, numerical simulation indicates $\delta_{c}\approx0.45$ \cite{Musco:2004ak,Musco:2008hv}. The threshold value of density contrast $\delta_{c}$ slightly decreases due to the softening of equation of state \cite{Musco:2012au,Saikawa:2018rcs} during the QCD phase transition \cite{Borsanyi:2016ksw,Saikawa:2018rcs}, and therefore the PBH mass function around $m\sim1\Msun$ should be significantly enhanced \cite{Byrnes:2018clq}. In this letter, the data of equation of state and the sound speed are adopted from \cite{Saikawa:2018rcs}.

Here, the best-fit values of $A$ and $\alpha$ are $A=0.004647$ and $\alpha=-0.1033$ for case I, and $A=0.004640$ and $\alpha=-0.1003$ for case II, and then the PBH mass function generated by such a broad tilted curvature power spectrum is illustrated by the blue (case I) and red (case II) curves in Fig.~\ref{fpbh_Pm_post}. In particular, the enhancement of PBH mass function around $m\sim1\Msun$ just attributes to the softening of the equation of state during the QCD phase transition. In a word, our results indicate that such a broad red-tilted curvature power spectrum provides a quite reasonable explanation for the PBH mass function implied by GWTC-3. 

{\it GWB associated with PBHs. } GWB is supposed to be a superposition of incoherent GWs that are characterized statistically. The GWB associated with PBHs includes two main contributions: one is the coalescences of binary PBHs \cite{Chen:2018rzo} (see more discussion in \cite{Mukherjee:2021ags,Mukherjee:2021itf}) and the other is SIGW inevitably generated by the curvature perturbations during the formation of PBHs \cite{Saito:2008jc}. 

\begin{figure*}[htbp!]
    \centering
    \includegraphics[width=\textwidth]{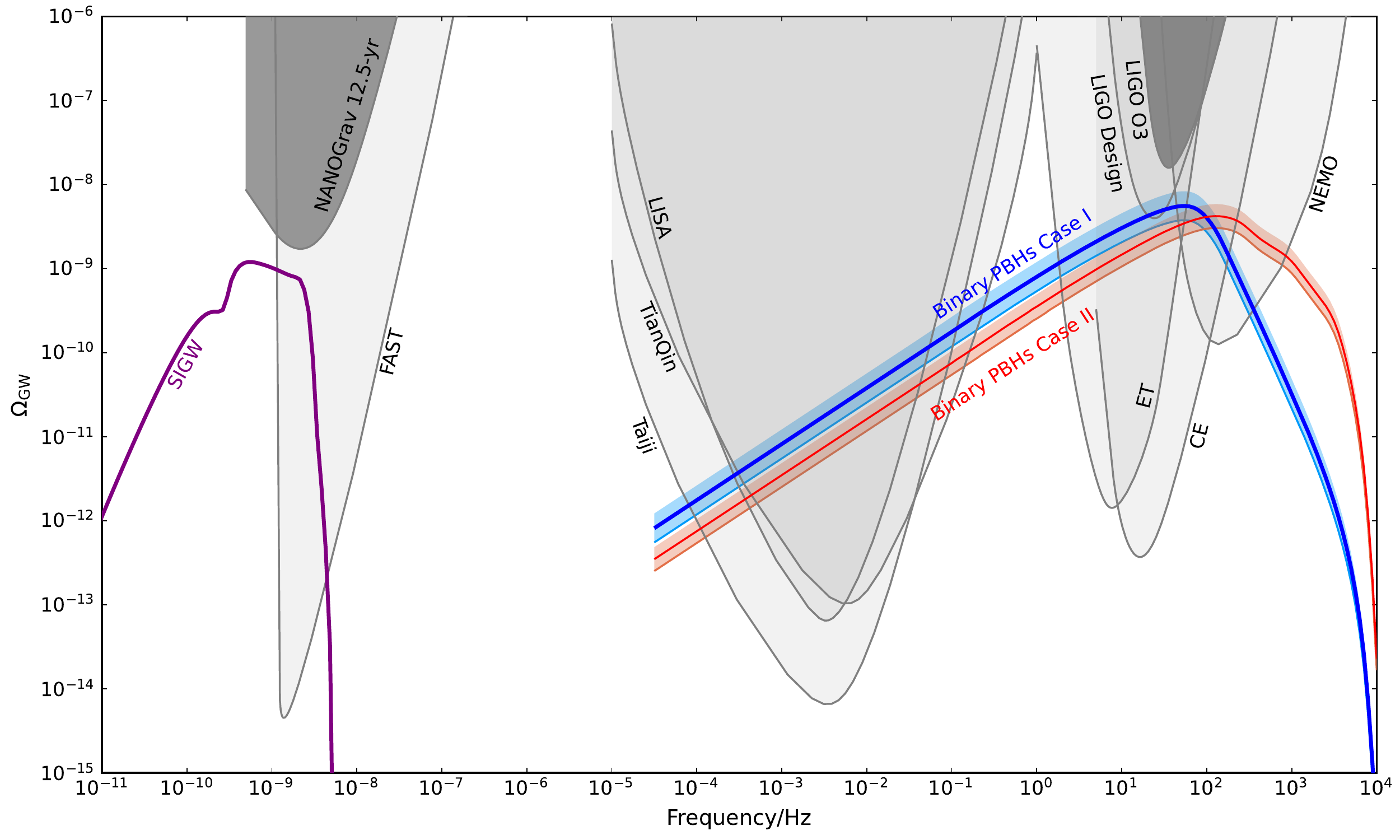}
    \caption{\label{ogw}
        The GWB associated with PBHs. The blue (red) band corresponds to the GWB from the coalescences of binary PBHs from Case I (Case II), and the purple curve corresponds to the SIGWs generated by the curvature perturbation during the formation of PBHs. The two dark shaded regions are ruled out by the LIGO O3 \cite{KAGRA:2021kbb} and NANOGrav 12.5-year data \cite{NANOGrav:2020bcs}, respectively. We also show the sensitivity curves of FAST \cite{Nan:2011um}, LISA \cite{Audley:2017drz}, TianQin \cite{TianQin:2015yph}, Taiji \cite{Hu:17}, LIGO Design, Einstein Telescope (ET) \cite{Punturo:2010zz}, Cosmic Explorer (CE) \cite{Evans:2016mbw} and NEMO \cite{Ackley:2020atn}.
    }
\end{figure*}

For binary systems, the energy-density spectrum of a GWB which defined as the energy density of GWs per logarithm frequency, $f=k/(2\pi)$, can be calculated as \cite{Allen:1997ad,Phinney:2001di,Regimbau:2008nj,Zhu:2011bd,Zhu:2012xw}
\e\label{OmegaGW}
\ogw(f) = \frac{f}{\rho_c H_0} \int \rd z \rd m_1 \rd m_2
\frac{\mR_{12}(z)}{(1+z)\, E(z)} \frac{\rd E_{\mathrm{GW}}}{\rd f_s},
\q
where $\rho_{c}=3H_{0}^2/(8 \pi)$ is the critical energy density of our Universe, $f_s$ is frequency of GWs in the source frame, $H_0$ is the Hubble constant,
and $E(z)=\sqrt{\Om_r \(1+z\)^4 + \Om_m (1+z)^3+\Omega_{\Lambda}}$ accounts for the evolution of our Universe with $\Omega_{\mathrm{r}}$, $\Omega_{\mathrm{m}}$ and $\Omega_{\mathrm{\Lambda}}$ the density parameters for radiation, matter and dark energy. Here, we adopt the best-fit results from Planck 2018 \citep{Planck:2018vyg} and approximate the energy spectrum $\rd E_{\mathrm{GW}}/\rd f_{s}$ emitted by an individual BBH using expressions from \cite{Cutler:1993vq,Chernoff:1993th,Zhu:2011bd}. The corresponding GWBs for case I and case II are respectively shown as the blue and red band  in \Fig{ogw}, indicating that both of them are compatible with the current limits given by LIGO O3 \cite{KAGRA:2021kbb}, and should be detected by the future ground-based and space-borne GW detectors, such as Neutron Star Extreme Matter Observatory (NEMO) \cite{Ackley:2020atn}, Cosmic Explorer (CE) \cite{Evans:2016mbw}, Einstein Telescope (ET) \cite{Punturo:2010zz}, Taiji \cite{Hu:17}, TianQin \cite{TianQin:2015yph} and LISA \cite{Audley:2017drz}.

On the other hand, it is known that the GWs should be generated by the scalar curvature perturbations at second order in perturbation theory \cite{tomita1967non,Matarrese:1992rp,Matarrese:1993zf,Matarrese:1997ay,Noh:2004bc,Carbone:2004iv,Nakamura:2004rm,Yuan:2019fwv,Yuan:2019udt,Yuan:2019wwo,Yuan:2021qgz}. In this sense, SIGWs are inevitably produced during the formation of PBHs. The perturbed metric in Newton gauge is given by 
\begin{equation}
	\mathrm{d} s^{2}=a^{2}\left\{-(1+2 \Phi) \mathrm{d} \eta^{2}+\left[(1-2 \Phi) \delta_{i j}+\frac{h_{i j}}{2}\right] \mathrm{d} x^{i} \mathrm{~d} x^{j}\right\},
\end{equation}
where $\Phi=-2\mathcal{R}/3$ is the Bardeen potential, $h_{ij}$ is the second-order transverse and traceless tensor mode and $\eta$ is the conformal time. The equation of motion for $\Phi$ is govern by 
\begin{equation}
	\Phi^{\prime \prime}+3 \mathcal{H}(\eta)\left(1+c_{s}^{2}\right) \Phi^{\prime}+3 \mathcal{H}^{2}\left(c_{s}^{2}-w\right) \Phi-c_{s}^{2} \nabla^{2} \Phi=0,
\end{equation}
where $w$ and $c_s$ are the equation of state and the sound speed, respectively. $\Phi_k(\eta)$ is related to its initial value $\Phi_{k}\equiv \Phi_{k}(\eta \rightarrow 0)$ by $\Phi_k(\eta)=\Phi_kT_\Phi(k\eta)$, where $T_\Phi(k\eta)$ is the transfer function. The equation of motion for the second-order tensor modes, $h_{ij}$, takes the form
\begin{equation}\label{eomh}
	h_{i j}^{\prime \prime}+2 \mathcal{H} h_{i j}^{\prime}-\nabla^{2} h_{i j}=-4 \mathcal{T}_{i j}^{\ell m} \mathcal{S}_{\ell m}
,
\end{equation}
with $\mathcal{T}_{i j}^{\ell m}=e_{i j}^{(+)}(\boldsymbol{k}) e^{(+) l m}(\boldsymbol{k})+e_{i j}^{(\times)}(\boldsymbol{k}) e^{(\times) l m}(\boldsymbol{k})$ selects the transverse-traceless part of the source term, with $e_{ij}$ the polarization tensor and $\mathcal{H}=a'/a$. The source term is given by
\begin{equation}
	S_{i j}=2 \Phi \partial_{i} \partial_{j} \Phi-\frac{4}{3(1+w)}\left(\partial_{i} \Phi+\frac{\partial_{i} \Phi^{\prime}}{\mathcal{H}(\eta)}\right)\left(\partial_{j} \Phi+\frac{\partial_{j} \Phi^{\prime}}{\mathcal{H}(\eta)}\right).
\end{equation}
Here the prime denotes the derivative with respect to $\eta$. Following \cite{Kohri:2018awv}, Eq.~(\ref{eomh}) can be solved by the Green's function and the transfer function method, and the energy density parameter by today is given by
\m
	\Omega_{\mathrm{GW}}&&=\frac{\Omega_{\mathrm{r}}}{6} \int_{0}^{\infty} \mathrm{d} u \int_{|1-u|}^{1+u} \mathrm{~d} v \frac{v^{2}}{u^{2}}\left[1-\left(\frac{1+v^{2}-u^{2}}{2 v}\right)^{2}\right]^{2}\nonumber\\
	&&
	\qquad\qquad
	\times \mathcal{P}_{\mathcal{R}}(u k) \mathcal{P}_{\mathcal{R}}(v k) \overline{I^{2}(u, v)},
\n
The kernel function takes the form \cite{Kohri:2018awv}
\m
&&\overline{I^2(u,v)}=\frac{9(u^2+v^2-3)^2}{32u^6v^6}\Bigg\{\Big(-4uv+(u^2+v^2-3)\nonumber\\
&&\times \ln\Big|{3-(u+v)^2\over 3-(u-v)^2}\Big|\Big)^2+\pi^2\(u^2+v^2-3\)^2\Theta(u+v-\sqrt{3})\Bigg\}.\nonumber\\
\n

For the curvature power spectrum given in Eq.~(\ref{pzeta}) with $A$ and $\alpha$ taking their best-fit values, the SIGWs are illustrated as the purple curve (coincides for case I and II) in \Fig{ogw}. We find that the predicted SIGWs are compatible with NANOGrav 12.5-yr data \cite{NANOGrav:2020bcs}, and should be detected by FAST \cite{Nan:2011um} in the future. Note that recent analysis implies there is no statistically significant evidence for the tensor transverse polarization mode in the NANOGrav 12.5-yr data set \cite{NANOGrav:2020bcs,Chen:2021wdo,NANOGrav:2021ini}, PPTA second data release \cite{Goncharov:2021oub,Wu:2021kmd}, IPTA second data release \cite{Chen:2021ncc}, and EPTA second data release \cite{Chalumeau:2021fpz}.  

{\it Conclusion and Discussion. } In this letter, we use a binned PBH mass function to model-independently reconstruct the PBH mass function from GWTC-3, and find that such a mass function can be naturally explained by a broad red-tilted curvature power spectrum. By comparing the merger rates of binary PBHs with those given by LVC, we conclude that GW190521 with the primary component being within the pair-instability supernovae mass gap can be explained by the merger of binary PBHs, but the light components (i.e. $m<3M_\odot$) in GW190814, GW190425, GW200105, and GW200115 events should be NSs or other exotic compact objects. In addition, the PBH scenario proposed in this letter can be tested by searching for the GWB generated by the binary PBHs and the SIGW inevitably produced by the curvature perturbations during the formation of PBHs.

{\it Acknowledgments. }
We acknowledge the use of HPC Cluster of ITP-CAS and HPC Cluster of Tianhe II in National Supercomputing Center in Guangzhou. This work is supported by the National Key Research and Development Program of China Grant No.2020YFC2201502, grants from NSFC (grant No. 11975019, 11991052, 12047503), Key Research Program of Frontier Sciences, CAS, Grant NO. ZDBS-LY-7009, CAS Project for Young Scientists in Basic Research YSBR-006, the Key Research Program of the Chinese Academy of Sciences (Grant NO. XDPB15), and the science research grants from the China Manned Space Project with NO. CMS-CSST-2021-B01.

\bibliography{./ref}
	
\end{document}